\documentclass[prl,%
reprint,
superscriptaddress,
 amsmath,amssymb,
 aps,
]{revtex4-2}

\usepackage{graphicx}
\usepackage{dcolumn}
\usepackage{bm}
\usepackage{babel}
\usepackage{appendix}
\usepackage{array}

\usepackage{tabularx}
\newcolumntype{L}[1]{>{\raggedright\arraybackslash}p{#1}}

\usepackage{xcolor}

\begin{document}

\preprint{APS/123-QED}

\title{Disentangling Lengthscales in Disordered Fracture}

\author{Matthaios~Chouzouris}%
\email{m.chouzouris@sms.ed.ac.uk}
\affiliation{School of Engineering, University of Edinburgh, Edinburgh EH9 3FG, UK}

\author{Marcelo~A.~Dias}%
\email{marcelo.dias@furman.edu}
\affiliation{School of Engineering, University of Edinburgh, Edinburgh EH9 3FG, UK}
\affiliation{Physics Department, Furman University, Greenville~29613, USA}

\makeatletter
\renewcommand\frontmatter@affiliationfont{\small}  
\makeatother

\date{\today}

\begin{abstract}
The interplay between quenched disorder and collective interactions can generate multiple emergent lengthscales in damage processes, yet their roles in determining the strength of heterogeneous materials remain incompletely understood. Here, we develop a coarse-grained description of load redistribution in the two-dimensional random fuse model (RFM) with Weibull-distributed quenched disorder, interpreting shielding in the crack-tip damage density as a consequence of nonlocal current redistribution. Our numerical results distinguish this emergent screening length, identified as the size of the statistical fracture process zone, from the regularizing length in Bažant’s established quasibrittle size-effect law (SEL), revealing their distinct disorder dependences, and challenging previous associations of the two. We further uncover a comparable scaling between the SEL and pre-failure crack extension, suggesting the former may instead be more naturally associated with a percolation process of micro-fractures around the crack-tip. These findings disentangle important emergent lengthscales within the RFM, and highlight a candidate physical mechanism for explaining macroscopic size effects in quasibrittle fracture.

\end{abstract}

\maketitle

The role of size in determining material strength represents a vital field of research, with open problems that have fascinated scientists and engineers for centuries. The field reaches back at least to Leonardo da Vinci's observation that longer wires sustain lower loads than shorter ones, and to Edme Mariotte's attribution of this effect to material inhomogeneities, or \textit{``the inequality of matter''} \cite{davinci1940meccanica, mariotte1718traite}. Weibull's theory later formalized these observations, under what is now known as the \textit{weakest-link} hypothesis: failure becomes more likely as a specimen samples more failure opportunities with the weakest point dictating its ultimate strength \cite{weibull_statistical_1939}. 

Weibull theory may be most simply stated at the level of a single microscopic element, subjected to a uniform load $\sigma$, where its survival probability follows:
\begin{equation}
 S(\sigma;\sigma_0,m)=\exp[-(\sigma/\sigma_0)^m],
\end{equation}
where $\sigma_0$ is a characteristic load scale and $m$ is known as the Weibull modulus. This model pertains to brittle fracture, introducing a size effect without an intrinsic material length: It maintains its functional form under renormalization group operations, in a scale-free manner \cite{bertin2010renormalization, shekhawat2013damage, chouzouris2026geometry}. However, it becomes incomplete when sufficient damage accumulates, with significant interactions and correlations arising before failure \cite{shekhawat2013damage}.

\begin{figure*}
\centering
\includegraphics[width=1.0\textwidth]{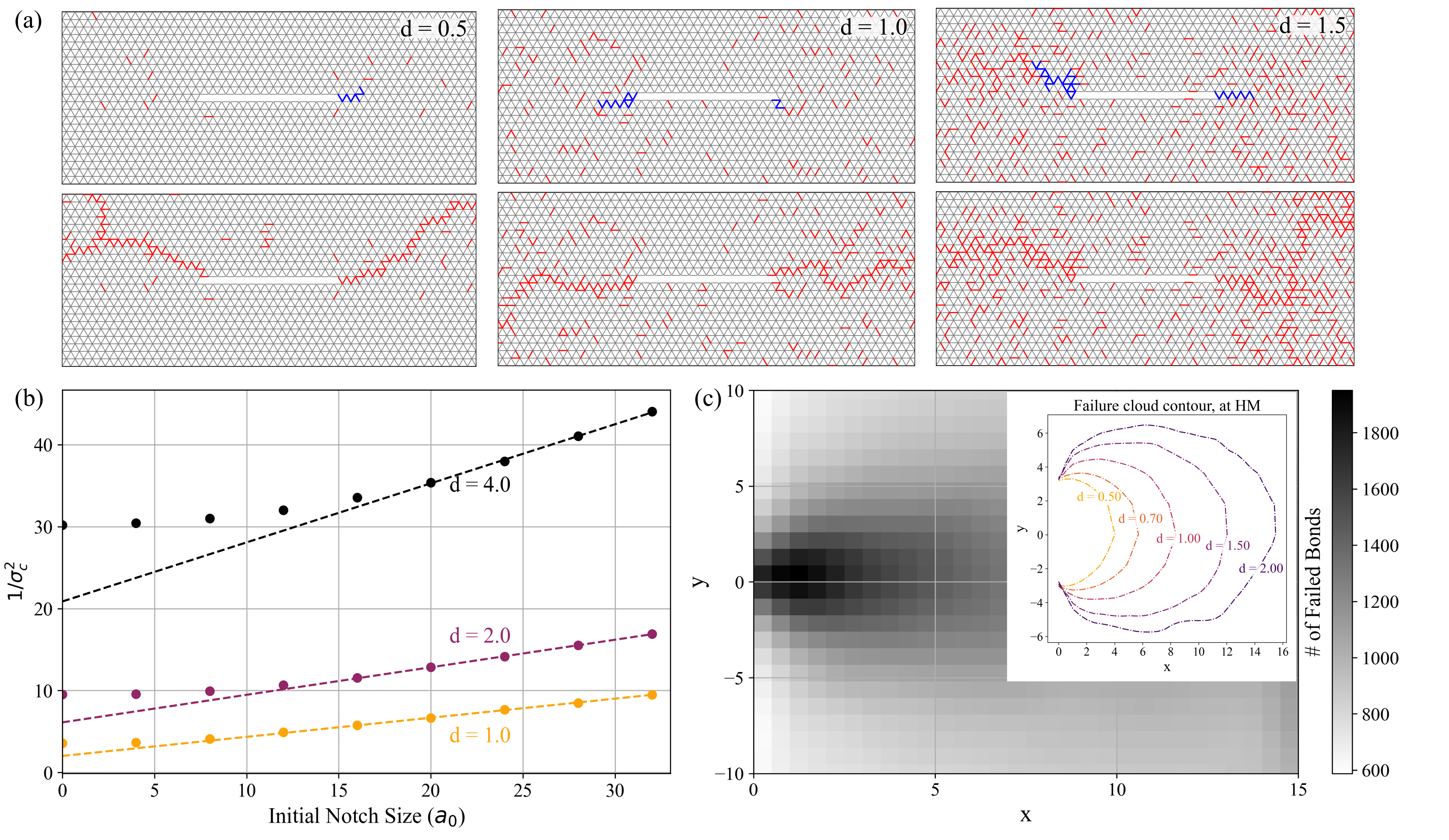}
\caption{Problem setup and size effect. (a) RFM damage patterns for different disorder strengths $d$. The pre-crack is visible at the center of each domain. Upper and lower panels show the peak-load and fully fractured configurations, respectively, from the same simulations. Gray and red elements denote intact and burnt fuses, respectively, while blue highlights the percolating cluster of burnt fuses advancing the pre-crack. Crack extension at peak load, $\Delta a$, is defined as the horizontal extent of these clusters. (b) Bažant's SEL: at large notch lengths, $1/\sigma_c^2\simeq(\xi+a_0)/K_c^2$, giving a linear dependence on $a_0$. For small notches, the data depart from the linear guides and the strength approaches a notch-length-independent plateau. (c) Spatial distribution of failed bonds at peak load for $d=1.0$, over $2000$ runs, with the crack tip at the origin. The inset shows the broadening of the half-maximum (HM) contour for this distribution, with increasing disorder.}
\label{fig:Figure1}
\end{figure*}

A central contribution to the understanding of such \textit{quasibrittle} failure, is Bažant's empirical size-effect law (SEL) \cite{bavzant1984size,bavzant1999size,bazant2019fracture}. This framework, captures the crossover between a regime where strength is size independent, to linear elastic fracture mechanics (LEFM), through a characteristic length $\xi$. Specifically and according to the SEL of interest, for a domain with a pre-crack of length $a_0$ the critical load $\sigma_c$ at which catastrophic failure is expected to occur, follows
\begin{equation}
 \sigma_c=\frac{K_c}{\sqrt{\xi+a_0 f(a_c/a_0)}},
\label{eq:SEL}
\end{equation}
where $K_c$ is a critical load scale. Here, $f$ mediates yet another crossover, hiding the pre-crack below a certain critical size $a_0 \lesssim a_c$ giving an effective $a_0 f(a_c/a_0) = a_c$, while on the other hand $f\to1$ for sufficiently large notches. In the latter, macroscopic crack regime, one retrieves the expected LEFM scaling for strength,
\begin{equation}
 \frac{1}{\sigma_c^2}\simeq\frac{1}{K_c^2}(\xi+a_0),
 \label{eq:SELlarge}
\end{equation}
so $\xi$ can be extracted from the macroscopic dependence of strength on notch size (Fig.~\ref{fig:Figure1}(b)). The lengthscale $\xi$ is often associated with the size of the fracture process zone (FPZ) \cite{alava2008fracture,bazant2019fracture}, a highly non-linear region developing around the crack-tip, yet the microscopic process setting this length remains contested.

The random fuse model (RFM) provides a minimal setting for fracture processes, allowing us to tackle this question while retaining quenched disorder and nonlocal load redistribution \cite{alava2008fracture, alava2008role, dearcangelis1985breaking, shekhawat2013damage, liu2026damage}. It involves quasistatically driving a two-dimensional simulated resistor network to failure, removing fuses when their currents exceed randomly assigned thresholds, with each removal redistributing current through the network, coupling local damage to its surroundings. We draw individual fuse failure thresholds from a Weibull distribution and quantify disorder by $d\equiv1/m$: increasing $d$ broadens the threshold distribution. For all subsequent measurements, data is extracted from RFM simulations of square lattice domains of side $L=128$ unit cells, notch sizes in $0\le a_0\le32$, and disorder strengths of $0.01\le d\le4$, with $2000$ runs for every pair of $a_0$ and $d$. In all cases a pre-cracked geometry involves a central notch placed parallel to the loading bus bars, as in previous RFM studies \cite{alava2008fracture,alava2008role,alava2006statistical}---see also Fig.\ref{fig:Figure1}(a). Changing $d$ then affects not only the morphology of the final fracture, but also the damage accumulated around the crack-tip at peak load---see Fig.~\ref{fig:Figure1}(a,c).

Previous RFM studies have connected $\xi$ to the spatial extent of an exponentially localized damage cloud surrounding the crack-tip, termed the statistical FPZ \cite{alava2008fracture, alava2008role}. This connection is intuitively appealing: just as $\xi$ effectively augments the initial notch size in the SEL Eq.~\eqref{eq:SEL}, the statistical FPZ appends itself to the crack (Fig.\ref{fig:Figure1}(c)). The exponentially decaying damage concentration around the crack-tip was conjectured to have a size scaling with disorder near $\sim d^{3/2}$ \cite{alava2008fracture}, using nonetheless power-law distributed quenched disorder and a limited range of disorder strengths---less than a decade. However, the LEFM load field has an algebraic, scale-free crack-tip singularity ($r^{-1/2}$), and the mechanism that screens the emerging damage profile has remained unclear. Moreover, a distinct length has recently been identified in the equiprobable damage zone (EDZ), defined as the region within which the damage density is approximately independent of distance from the tip \cite{liu2026damage}; its reported scaling of $\sim d^{4/3}$ consistent with the two-dimensional percolation exponent $\nu \equiv 4/3$ \cite{smirnov2001critical, hansen2003origin}. The relation between the SEL, the size of the statistical FPZ, and the scale of the EDZ therefore remains unclear.

Here we separate these lengths by independent measurements and analysis. We develop a coarse-grained model showing how nonlocal current redistribution may generate screening, with its mean-field limit providing an accurate parameter-free prediction for the nominal strength of un-notched samples. Through RFM simulations, we find that the size of the statistical FPZ follows a $\sim d^1$ scaling for a large portion of the domain of interest, consistent with simple theoretical considerations, whereas the SEL length scales with $\xi\sim d^{1.34(2)}$. Noting the connection with the EDZ scaling and the percolation critical exponent of $4/3$, we study clusters of connected microfractures extending the main crack, as a possible interpretation for the SEL scaling. Remarkably, we find that the mean crack extension at peak load (see Fig.\ref{fig:Figure1}(a)), grows as $\sim d^{1.33(3)}$, while also converging to the SEL length in magnitude for intermediate to large disorder. These results provide a microscopic interpretation of the length entering Bažant's quasibrittle size effect in Eq.~\eqref{eq:SEL} and may more generally clarify the distinct physical origins of the lengthscales governing fracture in quasibrittle materials.

\textit{Theoretical description.—} We begin by examining a coarse-grained description of the RFM. Let $\mathsf{T}(\bm r)$ denote the emergent load field in the network during the damage process. For Weibull disorder, the local density of broken bonds will then be expected to follow
\begin{equation}
 D(\bm r)=1-\exp\!\left[-\left(\frac{\mathsf{T}(\bm r)}{\mathsf{T}_0}\right)^{1/d}\right],
\end{equation}
where $\mathsf{T}_0$ sets the load scale. When $(\mathsf{T}/\mathsf{T}_0)^{1/d}\ll1$,
\begin{equation}
 D(\bm r)\simeq\left(\frac{\mathsf{T}(\bm r)}{\mathsf{T}_0}\right)^{1/d}.
 \label{eq:DfromJ}
\end{equation}
This approximation in conjunction with the scale-free $\sim r^{-1/2}$ solution for the crack-tip singularity in LEFM, would naively be expected to produce a scale-free damage concentration centred on the crack-tip, whose shape would inherit the power law dependence $\sim r^{-1/2d}$. In contrast, the observed emergent lengthscales and screening associated with the damage density around a crack \cite{alava2008fracture, alava2008role}, hint towards the importance of load redistribution because of damage.

Linear elasticity predicts that the change in the load field at $\boldsymbol{r}$, due to a bond breaking at $\boldsymbol{r}'$ decays as $\sim \mathsf{T}(\boldsymbol{r}') |\boldsymbol{r}' - \boldsymbol{r}|^{-2}$ (ignoring the dipolar directional dependence) \cite{bhagavatula1994green, shekhawat2013damage}. We represent this redistribution by a polar averaged kernel $\mathcal K(\bm r,\bm r')\sim|\bm r-\bm r'|^{-2}$. If $\sigma(\bm r)$ is the load field without damage, we write the non-local self-consistency condition for the emergent field as
\begin{align}
 \mathsf{T}(\bm r)&=\sigma(\bm r)+\Delta \mathsf{T}(\bm r),\\
 \Delta \mathsf{T}(\bm r)&=\int_\Omega\!d\bm r'\,\mathcal K(\bm r,\bm r')\mathsf{T}(\bm r')D(\bm r')\nonumber\\
 &= [\mathcal K*(D \mathsf{T})](\bm r).
 \label{eq:feedback}
\end{align}
Combining this relation with Eq.~\eqref{eq:DfromJ} yields
\begin{equation}
 D^d(\bm r)=\frac{\sigma(\bm r)}{\mathsf{T}_0}+[\mathcal K*D^{1+d}](\bm r).
 \label{eq:fullmodel}
\end{equation}

\begin{figure}
\includegraphics[width=\linewidth]{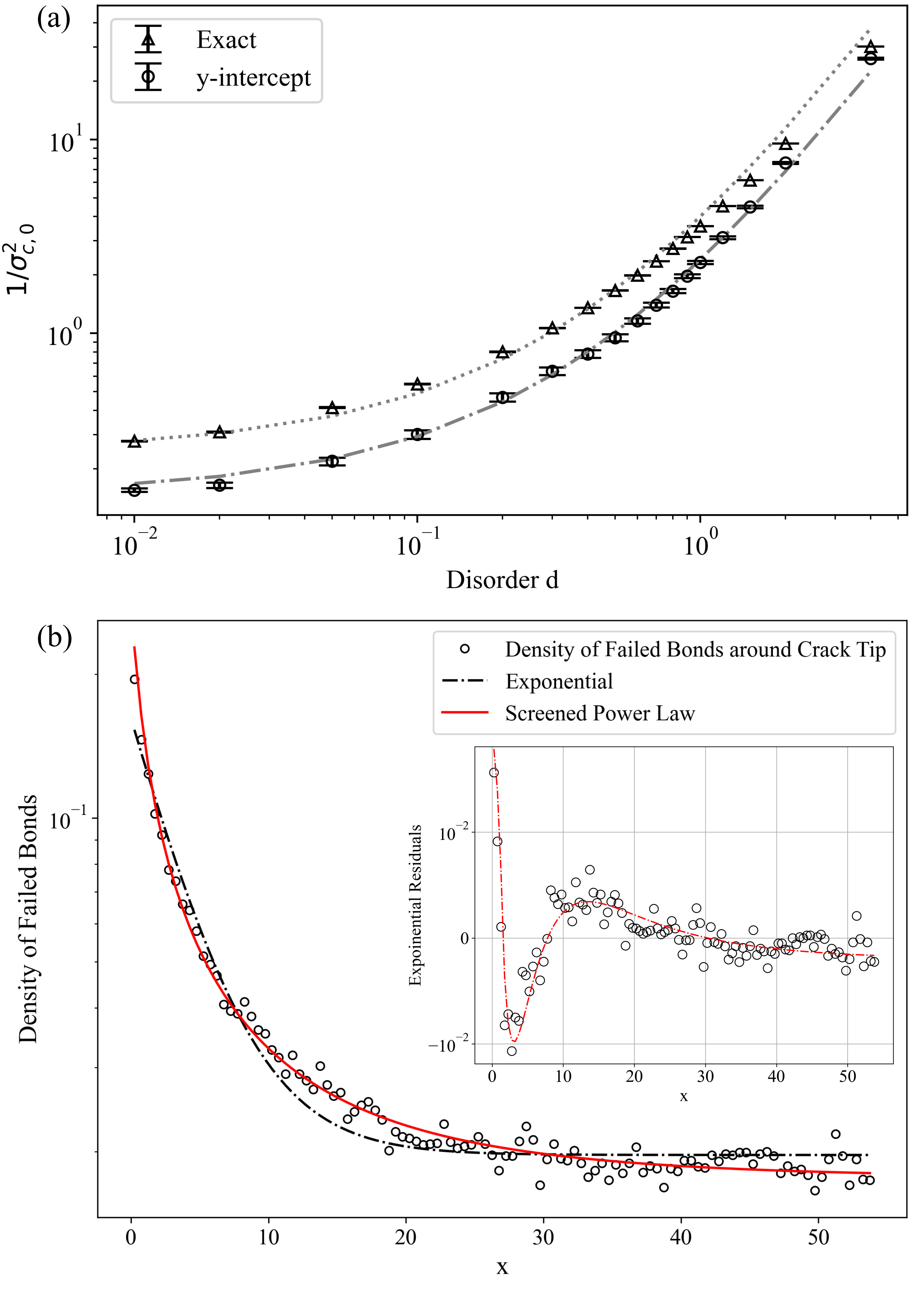}
\caption{\label{fig:Figure2}Mean-field strength and damage-cloud profile. (a) Critical load in un-notched samples versus disorder. The mean-field prediction Eq.~\eqref{eq:critical} (dotted line) has no free parameters; the strength extrapolated as the y-intercept of the large $a_0$ regime Eq.~\eqref{eq:SELlarge}, follows a similar trend after fitting a single pre-factor (dashed line). (b) Damage density projected along the notch axis for $d = 1.0$. The screened power-law form in Eq.~\eqref{eq:Dprofile} and a pure exponential are fitted to the profile; the inset shows the residuals of the exponential fit and the improvement achieved by Eq.~\eqref{eq:Dprofile}.}
\end{figure}

To obtain a local long-wavelength approximation, we expand the convolution in spatial gradients, in terms of kernel moments $\{ \mathcal{K}_i \}$: $\mathcal K*g=\mathcal K_0g+(\mathcal K_2/2)\nabla^2g+\cdots$. Odd moments vanish by symmetry, even with the angular structure considered. As the kernel is marginal in 2D, this expansion requires a finite domain $L$ and a lattice-scale cutoff $\alpha$, at the kernel origin; accordingly, the coefficient $\mathcal K_2 \sim (L^2 - \alpha^2)$, should be regularized by these system size dependent cutoffs, rather than assumed to be a universal material constant. Normalizing the zeroth moment $\mathcal K_0$ to unity, consistent with current conservation in the finite network, locally gives $\mathsf{T}(\bm r) = \sigma(\bm r)(1 - D(\bm r))^{-1}$; a \textit{stiffness reduction factoring} \cite{pijaudier1987nonlocal, pradhan2010failure}. Thus:
\begin{equation}
 [1-D(\bm r)]D^d(\bm r)-\frac{\sigma(\bm r)}{\mathsf{T}_0}
 =\frac{\mathcal K_2}{2}\nabla^2D^{1+d}(\bm r).
 \label{eq:localPDE}
\end{equation}
This equation is an approximate coarse-grained closure for the RFM \footnote{Note that the local load balance of Eq.~\eqref{eq:localPDE} is equivalent to a fiber bundle model with global load sharing \cite{pradhan2010failure}.}. Its two useful limits are a uniform mean-field prediction for the strength of un-notched specimens, and a linear-response description of screening, which we study below.

\textit{Mean-field strength.—} For uniform damage, the gradient term in Eq.~\eqref{eq:localPDE} vanishes and $\sigma=\mathsf{T}_0D^d(1-D)$. Then the maximum sustainable load for an initially un-notched system occurs at
\begin{equation}
 D_c=\frac{d}{1+d},\qquad
 \sigma_{c,0}=\mathsf{T}_0(1+d)^{-1}(1+1/d)^{-d}.
 \label{eq:critical}
\end{equation}
For our network topology, $\mathsf{T}_0=2$. Fig.~\ref{fig:Figure2}(a) compares Eq.~\eqref{eq:critical} with the measured critical current of un-notched samples. Despite the elementary mean-field approximations, the parameter-free prediction captures the trend well for over more than two decades in $d$, extending beyond the stated range of validity of the approximation. The strength extrapolated from the LEFM mediated regime follows the same dependence up to one fitted multiplicative factor. This agreement does not validate the full spatial theory, but gives a useful check on its local balance and stability.

\textit{Screening of the damage cloud.—} The feedback in Eq.~\eqref{eq:fullmodel} generates a characteristic length in the linear response about a uniform background. Writing $\mathsf{T}=\mathsf{T}_B+\delta \mathsf{T}$ and $\sigma=\sigma_B+\delta \sigma$, with $D_B=(\mathsf{T}_B/\mathsf{T}_0)^{1/d}$, gives a modified Helmholtz equation
\begin{equation}
 [1-\xi_{\mathsf{T}}^2\nabla^2]\delta \mathsf{T}(\bm r)
 =\frac{\delta \sigma(\bm r)}{1-D_B/D_c},
 \label{eq:helmholtz}
\end{equation}
where
\begin{equation}
 \xi_{\mathsf{T}}=\sqrt{\frac{\mathcal K_2 D_B}{2(D_c-D_B)}}.
 \label{eq:xiJ}
\end{equation}
This linear-response length diverges as $D_B\to D_c$, signaling the loss of stability of the uniform mean-field state. Moreover, the system-size regularization of $\mathcal{K}_2$ gives $\mathcal{K}_2\sim L^2$, and hence predicts $\xi_{\mathsf{T}}\sim L$, provided $D_B$ and $D_c$ remain weakly dependent on $L$. This scaling is particularly interesting in the context of crack roughness. Previous studies have found that the extent of the damage zone surrounding the main crack scales linearly with the system size \cite{hansen2003origin}. This same scaling is a necessary ingredient in theoretical estimates of crack-roughness exponents, which are consistent with numerical results for the RFM \cite{hansen2003origin, nukala2008anomalous} and experimental measurements in silica glass and aluminum alloys \cite{ponson2006two}. The predicted $\xi_{\mathsf{T}}\sim L$ is thus consistent with the system-size scaling underlying these theoretical descriptions of crack roughness.

Regarding the functional form of the damage zone profile, the homogeneous response in two dimensions contains a Bessel function $K_0(r/\xi_{\mathsf{T}})\sim e^{-r/\xi_{\mathsf{T}}}r^{-1/2}$ at distances $r\gtrsim\xi_{\mathsf{T}}$. Thus, it establishes a possible mechanism for the observed screening through nonlocal load redistribution; it does not predict the exact disorder dependence or form of the damage cloud around a macroscopic notch, which would require solving the full non-linear problem. 

The profile also retains an algebraic contribution: for weak gradients and LEFM forcing $\sigma(r)\sim r^{-1/2}$, Eq.~\eqref{eq:localPDE} recovers $D(r)\sim r^{-1/(2d)}$. We therefore fit the measured damage density with the phenomenological crossover form
\begin{equation}
 D(r)\simeq A\frac{e^{-r/\xi_{\rm scr}}}{(r+1)^{1/(2d)}}+D_B ,
 \label{eq:Dprofile}
\end{equation}
regularizing the denominator with the lattice spacing $\alpha$ (unity). This form combines the algebraic modulation expected from Weibull statistics with an effective exponential cutoff motivated by the screened response. It is an ansatz for the nonlinear profile---not an exact superposition of solutions---and its fitted length $\xi_{\rm scr}$ need not equal the linear-response length $\xi_{\mathsf{T}}$. As Fig.~\ref{fig:Figure2}(b) shows, it captures the observed profile better than a pure exponential for the largest range of $d$ values, with the same number of fit parameters, explaining fitting residuals observed in previous studies \cite{alava2008fracture}. A more complete discussion of the improvement over all $d$ can be found in the SI.

A simple conditional argument can actually motivate a linear dependence of the screening lengthscales $\xi_{exp}$ and $\xi_{scr}$ on $d$, within a certain approximation. For low enough disorder, we expect limited damage and current redistribution, and thus $\mathsf{T}(\boldsymbol{r}) \approx \sigma(\boldsymbol{r})$. Then any regularized load profile can in general be approximated around the crack-tip by $\mathsf{T}(r)= a(1-br) + \mathcal{O}(r^2)$, with $a$ and $b$ independent of disorder, and through Eq.~\eqref{eq:DfromJ}
\begin{equation}
 D(r) = \left(\frac{a}{\mathsf{T}_0}\right)^{1/d}
 \left(1-\frac{b}{d}r\right) + \mathcal{O}(r^2).
\end{equation}
The approximate decay rate of this profile to lowest order is then proportional to $1/d$, suggesting a characteristic length $\sim d^1$. This argument is a rationale under stated assumptions, and particularly Eq.~\eqref{eq:DfromJ}, and would be equivalent for power-law distributed quenched disorder, such as in \cite{alava2008fracture, alava2008role}, and in contrast to the previously conjectured $\sim d^{3/2}$ scaling of $\xi_{exp}$ \cite{alava2008fracture}, which was nonetheless neither theoretically motivated nor numerically supported. Our conditional argument does not represent a rigorous derivation of the full disorder dependence either and is expected to fail beyond a certain value of $d$; the actual scaling is now established from simulations.

\begin{figure}
\includegraphics[width=\linewidth]{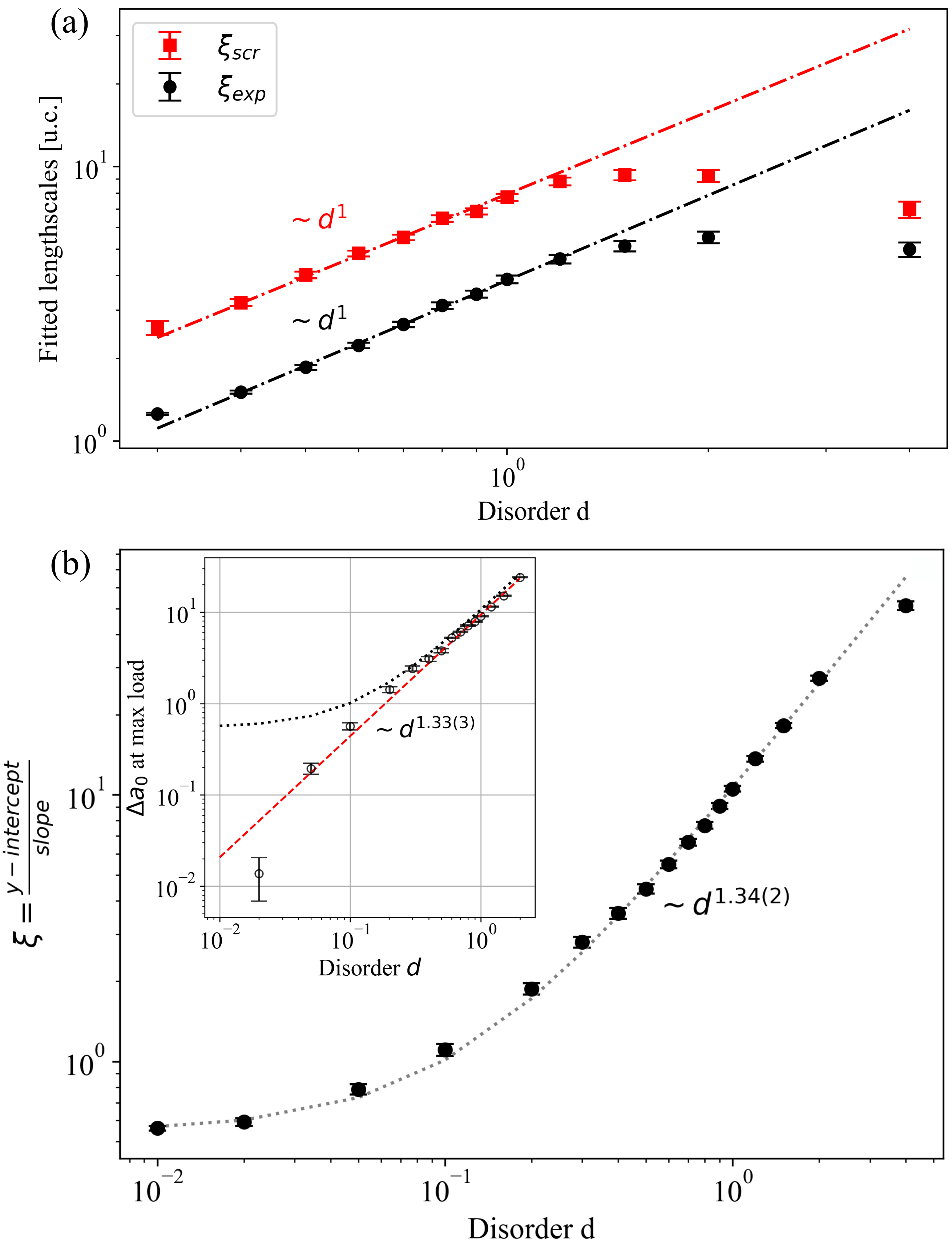}
\caption{\label{fig:Figure3}Measured fracture lengthscales. (a) Lengths extracted from the screened power-law and pure-exponential fits to damage profiles. Both show an approximately $d^1$ regime; reliable estimates are unavailable below $d=0.3$, and departures from the guides occur at higher disorder. (b) SEL length $\xi$ fitted by $\xi(d)=\xi_0+cd^\beta$, with $\beta=1.34(2)$ (black dotted line). The inset shows the average horizontal extension of the original notch at peak load, a power-law fit scaling $\Delta a_0\sim d^{1.33(3)}$ (red dashed line), as well as the SEL fit $\xi(d)$ from the main plot (black dotted line).}
\end{figure}

\textit{Distinct disorder scalings.—} Here we compare the independently measured lengthscales of interest. Following our simple scaling argument, the fitted screening lengths $\xi_{\rm scr}$ and $\xi_{\exp}$ obtained from fits to the damage cloud, both exhibit an approximately linear dependence on $d$ over a substantial intermediate to large disorder range, as highlighted in Fig.~\ref{fig:Figure3}(a). At very large disorder, the fitted lengths depart from the linear guides. Importantly, the profile differs from the $d^{3/2}$ dependence conjectured in earlier work, across the whole measurement domain.

The SEL length behaves differently. Anticipating a stress intensity regularization due to lattice spacing \cite{nguyen2019role}, we fit our measurements to a regularized power-law $\xi=\xi_0+cd^\beta$ with free parameters $\{\xi_0, c, \beta\}$, giving good agreement with measurements over more than two decades, and an extracted $\beta=1.34(2)$ across the range shown in Fig.~\ref{fig:Figure3}(b). This is numerically consistent with $4/3$, the two-dimensional percolation correlation-length exponent, and the scaling reported for the EDZ \cite{liu2026damage}. Notably, it is also consistent with a previously proposed classification of the RFM within the percolation universality class \cite{hansen2003origin}, as well as with the disorder scaling of the percolation correlation length at peak load, in a mean-field approximation \footnote{Using the dilution factor $D$ as the control parameter, and given the mean-field critical point $D^* = 1$, the system reaches peak load at a distance $\epsilon \equiv |D_c - D^*| = 1/(1+d)$ from the critical point. For large $d$, the percolation correlation length is then $\xi_{perc}\sim\epsilon^{-\nu}\sim d^{\nu}$.}. Importantly, it differs from the $\sim d^1$ regime of both fitted damage-cloud lengths and their overall disorder dependence. We conclude that, within the RFM, the SEL length cannot be simply identified with the spatial extent of the statistical FPZ, in clear contrast to previous findings \cite{alava2008role,alava2008fracture}.

What, then, does $\xi$ physically represent? The scaling association with the EDZ and two-dimensional percolation, suggests a potential link with the size of connected micro-fracture clusters within the damage cloud, that could be effectively extending the pre-crack before failure. We measure and analyze the average horizontal extension of the central notch $\Delta a_0$ at peak load, due to clusters of burnt fuses connected to it (blue elements in Fig.\ref{fig:Figure1}(a)). As shown in the inset of Fig.~\ref{fig:Figure3}(b), this extension scales closely with $\Delta a_0 \sim d^{1.33(3)}$ over almost two decades, matching $\xi$ not only in exponent, but also converging in magnitude from intermediate to large disorder. Although this observation does not establish a causal connection, it suggests that macroscopic size effects may more naturally reflect the growth of connected micro-fracture domains, rather than the size of the statistical FPZ or damage cloud.

\textit{Discussion.—} We have demonstrated that the lengthscales pertaining damage localization and macroscopic size effects in statistical fracture have distinct disorder dependences. While the size of the statistical FPZ grows linearly with disorder, the length related to the quasibrittle SEL follows a regularized power-law with large disorder limit $\xi\sim d^{1.34(2)}$, consistent with the two-dimensional percolation exponent related to the EDZ. The matching scaling of the pre-failure crack extension at peak load, $\Delta a_0\sim d^{1.33(3)}$, and its convergence in magnitude to the SEL lengthscale, suggests instead that the macroscopic size effect may be more naturally associated with the growth of a percolating cluster of burnt fuses at peak load, rather than the spatial extent of the statistical FPZ. Our coarse-grained analysis provides a mechanism for the emergence of screening through nonlocal current redistribution, offering a physical interpretation of the exponentially screened damage statistics and the linear disorder scaling observed in simulations. Together, these findings distinguish the spatial extent of damage and the statistical FPZ, from the topological organization of micro-fractures within it, and are suggestive of a natural connection of quasibrittle size effects with the latter.

Several questions remain open. Although the coarse-grained theory accounts for the emergence of screening, it does not trivially predict the exact form of the damage profile, or the full disorder scaling of the statistical FPZ. Moreover, although the agreement between the SEL length and crack extension is intuitively appealing and supports a causal connection between fracture connectivity at peak load and macroscopic size effects, it does not rigorously establish it, nor do the exponents alone establish that the relevant clusters belong to the percolation universality class. In that regard, we suggest that a rigorous theoretical understanding of the emergence of the EDZ, could help determine the microscopic origin of this relationship. Finally, whether the distinct scaling laws identified here persist across different forms of quenched disorder distributions and beyond the RFM remains an open question. Resolving these issues could establish the generality of the separation between coarse grained damage localization and topological micro-fracture connectivity, solidifying the physical meaning of the characteristic length governing quasibrittle fracture.

\textit{Availability.— }
The simulation code and datasets supporting the findings of this article are available at \cite{RFMRepo} and \cite{chouzouris_2026_24198980} respectively.

\begin{acknowledgments}
The authors thank UKRI for support under the EPSRC Open Fellowship scheme (Project No. EP/W019450/1).
\end{acknowledgments}


\bibliography{bibliography}

\clearpage
\newpage
\appendix

\section{Appendix}
In this supplementary information, we provide a nomenclature section, an overview of the simulations details, a goodness of fit study for the two damage density models we use in the main script, as well as a more extensive discussion on the approach to the percolation critical point in the mean-field limit, and its dependence on disorder strength.\\

\noindent\textbf{Nomenclature}

\medskip
\textit{Symbols}

$\sigma, \mathsf{T}$ -- Applied and emergent load/stress

$\sigma_0, \mathsf{T}_0$ -- Load scale in Weibull law

$m$ -- Weibull modulus

$d$ -- Disorder strength, $1/m$

$K_c$ -- Nominal stress intensity scale

$a_0$ -- Pre-crack (notch) size

$a_c$ -- Critical notch size

$f$ -- Crossover function

$\xi$ -- Bazant's SEL lengthscale

$\xi_{\mathrm{exp}}$ -- Exponential model fitted lengthscale

$\xi_{\mathrm{scr}}$ -- Screened power-law fitted lengthscale

$\xi_T$ -- Linearized continuum lengthscale

$\mathcal{K}, \mathcal{K}_i$ -- Load redistribution kernel and its $i$th moment

$L$ -- System size

$\alpha$ -- Lattice scale cutoff

$D_B, D_c$ -- Background and critical damage density

$\nu$ -- Percolation critical exponent

\medskip
\textit{Abbreviations}

SEL -- Size Effect Law

LEFM -- Linear Elastic Fracture Mechanics

FPZ -- Fracture Process Zone

RFM -- Random Fuse Model

EDZ -- Equiprobable Damage Zone
\medskip

\textbf{Quasistatic Random Fuse Simulator.- }
The simulator implements a two-dimensional random-fuse model on an $L\times L$ triangular lattice. Each intact bond $j$, has unit conductance and a quenched failure threshold $\mathsf T_{c}^{(j)}$, sampled from a Weibull distribution. For a given network configuration, the node voltages satisfy Kirchhoff’s equations,
\begin{equation}
    A\mathbf V=0,
\end{equation}
where $A$ is the sparse graph Laplacian of the network, and $\mathbf{V}$ is a vector that holds the voltages on all nodes. Dirichlet conditions $V=0$ and $V=V_{\mathrm{app}}$ are imposed on the bottom and top rows (the bus bars), respectively, and the reduced system for the interior nodes is solved using a conjugate-gradient method \cite{shewchuk1994introduction}. Periodic boundary conditions are maintained on the vertical edges, as in \cite{alava2008role, alava2008fracture}.

The current in bond $j$, connecting nodes $p$ and $q$, is
\begin{equation}
    \mathsf T^{(j)}=|V_p-V_q|,
\end{equation}
and the bond fails irreversibly when $\mathsf T^{(j)}>\mathsf T_{c}^{(j)}$. The applied voltage is increased quasi-statically; a bisection procedure refines the voltage increment until the next failure is isolated. Following each primary failure, the graph Laplacian is updated by removing the corresponding bond, the voltage field is recomputed, and any newly overstressed bonds are removed sequentially. This process may produce fracture avalanches through current redistribution.

\textbf{Statistical FPZ and Fitted Models.- }In the main text, we have used two distinct models---both with three free parameters---, to fit the shape of the statistical FPZ. These involved our phenomenologically motivated screened power-law,
\begin{equation}\label{eq:screened_power}
    f_1(r) = A\frac{e^{-r/\xi_{\rm scr}}}{(r+1)^{1/(2d)}}+D_B,
\end{equation}
and the exponential law introduced in \cite{alava2008fracture, alava2008role}:
\begin{equation}\label{eq:pure_exponential}
    f_2(r) = A' \ e^{-r/\xi_{exp}} + D'_B.
\end{equation}
In the main text, we claimed---without full validation---that the screened power-law produces better fits to the underlying damage density distribution. Here, we compare a goodness-of-fit metric for these two models through the range of disorders $d$. Specifically, we use the residual sum of squares (RSS), defined for a set of $N$ data-points $\{r_i, d_i\}$, as
\begin{equation}
    \textit{RSS} = \sum_{i=1}^N (d_i - f(r_i))^2
\end{equation}

\begin{figure}
\centering
\includegraphics[width=1.0\linewidth]{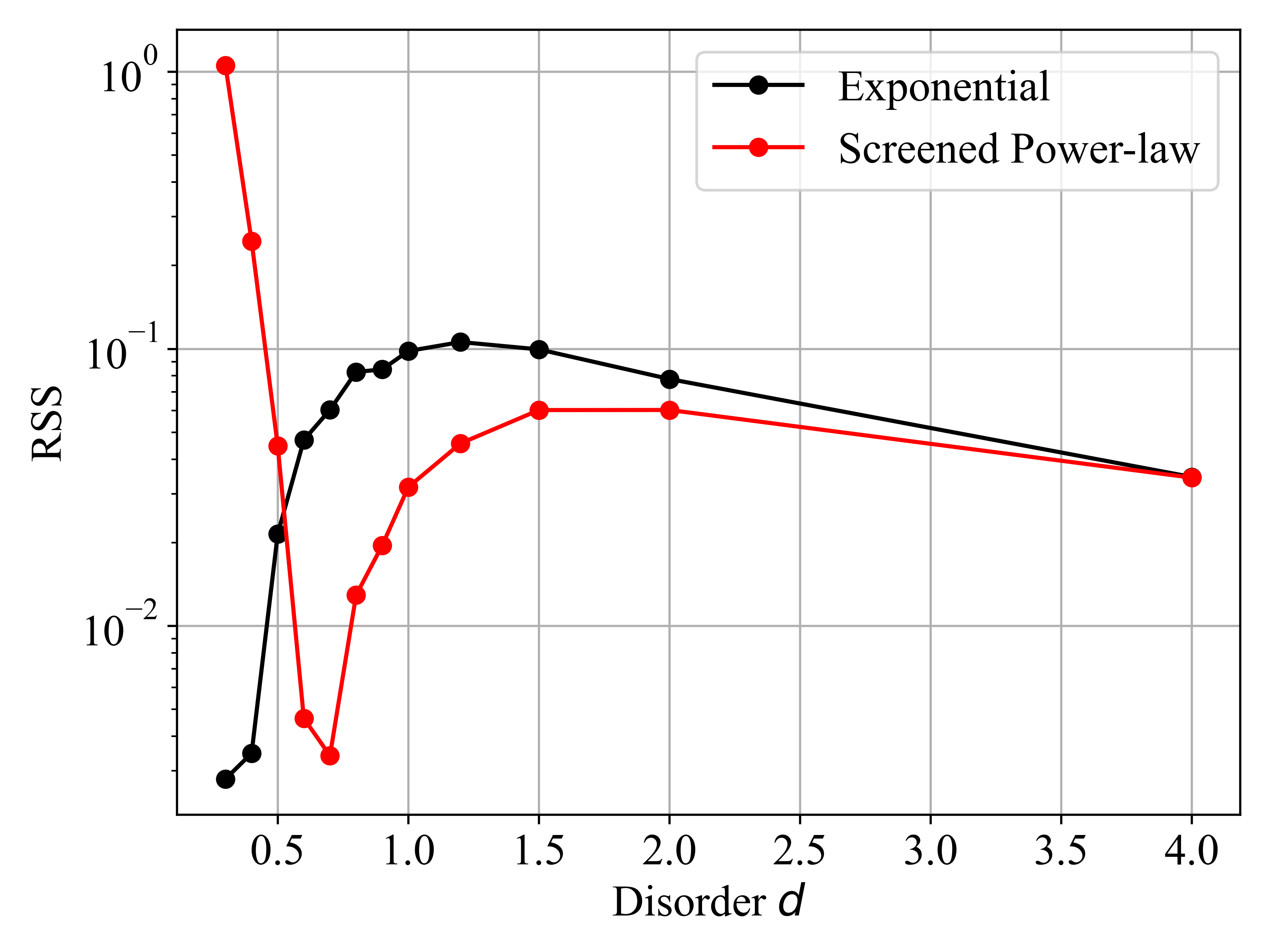}
\caption{Residual Sum of Squares (RSS), for the two models used to capture the shape of the damage cloud in the main script: the screened power-law (Eq.\ref{eq:screened_power}) offers an improvement over the pure exponential (Eq.\ref{eq:pure_exponential}) for the majority of the domain of $d$.}
\label{fig:SIFigure1}
\end{figure}

As depicted in Fig.\ref{fig:SIFigure1} the screened power-law has a lower RSS for most disorder values. However, for low disorder $d\leq0.4$, the two reverse with the screened power-law failing to capture the observed behaviour as well. We conjecture that for low $d$, the large majority of the damaged fuses close to the crack-tip is a result of pre-failure crack advance, without enough diffuse damage to sample the underlying stress distribution and thus give rise to the $\sim r^{-1/2d}$ damage density dependence close to the tip. For the rest of the domain the improvement is noticeable, reducing the RSS by a $0.5-1$ decades, for most of the relevant $d$ range, with the two eventually converging at large disorder values.

\textbf{Mean-field and Percolation.- }As the lattice incurs damage, and a proportion $D$ of bonds is removed from some region, the topology changes, introducing gaps; we briefly discuss the effect of this thinning on the possible emergence of the percolation exponent and associated lengthscales. Specifically, we consider high enough disorder $d$, so that the lattice reaches a level of thinning that brings it close to the percolation critical point $D^*$, with an emergent percolation lengthscale $\xi_{perc} \sim |D-D^*|^{-\nu}$. Although this discussion is significantly simplified, we briefly discuss how the distance from criticality at failure relates to the underlying disorder $|D-D^*|\sim d^{-\gamma}$, in the context of the mean-field limit of the local damage PDE. The mean-field limit is equivalent to the very extensively studied Fiber Bundle Model (FBM) with global load sharing \cite{pradhan2010failure}---a mechanical model for a collection of fibers sharing an applied load $\sigma$.

As discussed in the main text, in mean-field the emergent load $\mathsf{T}$ in terms of the global applied load $\sigma$, follows from a stiffness reduction factoring:
\begin{equation}
    \mathsf{T} = \frac{\sigma}{1 - D},
\end{equation}
leading to a dilution factor at peak load of
\begin{equation}
    D_{c} = \frac{d}{1+d}.
\end{equation}
Defining the distance from the critical point transition point ($D^* = 1$ in mean-field and the FBM), $\epsilon \equiv |D_c - D^*| = 1/(1+d)$, which for large $d$ yields a percolation lengthscale at peak load of:
\begin{equation}
    \xi_{perc} \sim \epsilon^{-\nu} \sim d^{\nu}.
\end{equation}
This is not necessarily consistent with the full picture when positional correlations appear in the damage field due to load redistribution. We however present it as a simple argument for the appearance of the percolation lengthscale critical exponent within a sufficiently damaged region, such as the EDZ. In \cite{liu2026damage}, the authors also note that the EDZ $\sim d^{4/3}$ scaling, breaks down at low enough disorder; it is still an open problem to determine how the zone arises in the first place, as well as discuss that subsequent scaling and regularization for more brittle materials.


\end{document}